\documentclass[superscriptaddress,prl,twocolumn,amssymb,nofootinbib,aps]{revtex4-1}
\usepackage{graphicx,bm,times}
\usepackage{tabularx} 
\usepackage{lipsum}
\usepackage{amsmath}
\usepackage{gensymb}
\usepackage{comment}
\usepackage{color,soul, mathrsfs}
\usepackage[utf8]{inputenc}
\usepackage[english]{babel}
\usepackage{natbib}
\usepackage{mathtools}
\usepackage{float}
\usepackage{physics}
\usepackage{siunitx}
\usepackage{hyperref}
\hypersetup{colorlinks=true,linkcolor=blue,citecolor=blue,urlcolor=black}
\usepackage{amsmath, amsthm, amssymb}
\usepackage{subfigure}
\usepackage{url}
\usepackage{tikz}
\usepackage{gensymb}
\usepackage{bbold}

\DeclareMathOperator{\Var}{Var}

\newcommand{\R}{\mathfrak{R}}
\newcommand{\A}{\hat{a}}
\newcommand{\V}{\nu}
\newcommand{\VB}{\overline{\nu}}
\newcommand{\VBB}{\overline{\overline{\nu}}}
\newcommand{\VBBB}{\overline{\overline{\overline{\nu}}}}

\newcommand{\M}{\mu}
\newcommand{\MB}{\overline{\mu}}
\newcommand{\MBB}{\overline{\overline{\mu}}}
\newcommand{\MBBB}{\overline{\overline{\overline{\mu}}}}
\newcommand{\la}{\langle}
\newcommand{\ra}{\rangle}
\newcommand{\x}{\hat{x}_\theta}
\newcommand{\D}{^\dagger}
\newcommand{\PZ}{\Psi_0}
\newcommand{\PM}{\Psi_m}
\newcommand{\AL}{\alpha}
\newcommand{\OM}{\Omega}
\newcommand{\n}{n_e}
\allowdisplaybreaks[3]

\begin{document}

\title{Quantum Enhanced Precision Estimation of Transmission with Bright Squeezed Light\\}

\author{G. S. Atkinson}
\email{george.atkinson@bristol.ac.uk}
\affiliation{Quantum Engineering Technology Labs, H. H. Wills Physics Laboratory and Department of Electrical \& Electronic Engineering, University of Bristol, Tyndall Avenue, BS8 1FD, United Kingdom.}
\affiliation{Quantum Engineering Centre for Doctoral Training, H. H. Wills Physics Laboratory and Department of Electrical \& Electronic Engineering, University of Bristol, Tyndall Avenue, BS8 1FD, United Kingdom.}
\author{E. J. Allen}
\email{euan.allen@bristol.ac.uk}
\affiliation{Quantum Engineering Technology Labs, H. H. Wills Physics Laboratory and Department of Electrical \& Electronic Engineering, University of Bristol, Tyndall Avenue, BS8 1FD, United Kingdom.}
\author{G. Ferranti}
\affiliation{Quantum Engineering Technology Labs, H. H. Wills Physics Laboratory and Department of Electrical \& Electronic Engineering, University of Bristol, Tyndall Avenue, BS8 1FD, United Kingdom.}
\author{A. R. McMillan}
\affiliation{Quantum Engineering Technology Labs, H. H. Wills Physics Laboratory and Department of Electrical \& Electronic Engineering, University of Bristol, Tyndall Avenue, BS8 1FD, United Kingdom.}
\author{J. C. F. Matthews}
\email{jonathan.matthews@bristol.ac.uk}
\affiliation{Quantum Engineering Technology Labs, H. H. Wills Physics Laboratory and Department of Electrical \& Electronic Engineering, University of Bristol, Tyndall Avenue, BS8 1FD, United Kingdom.}

\date{\today}

\begin{abstract} 
Squeezed light enables measurements with sensitivity beyond the quantum noise limit (QNL) for optical techniques such as spectroscopy, gravitational wave detection, magnetometry and imaging. Precision of a measurement — as quantified by the variance of repeated estimates — has also been enhanced beyond the QNL using squeezed light. However, sub-QNL sensitivity is not sufficient to achieve sub-QNL precision. Furthermore, demonstrations of sub-QNL precision in estimating transmission have been limited to picowatts of probe power. Here we demonstrate simultaneous enhancement of precision and sensitivity to beyond the QNL for estimating modulated transmission with a squeezed amplitude probe of $0.2$~mW average ($25$~W peak) power, which is 8 orders of magnitude above the power limitations of previous sub-QNL precision measurements of transmission. Our approach enables measurements that compete with the optical powers of current classical techniques, but have both improved precision and sensitivity beyond the classical limit.
\end{abstract}

\maketitle

Optical measurements are fundamentally limited by quantum fluctuations in the probe. The Poisson distributed photon number $n$ of coherent light -- often used as a probe in classical experiments -- results in shot-noise, which represents the quantum noise limit (QNL) in the precision of parameter estimation with classical resources~\cite{bachor2004guide}. Because the QNL scales with $\sim1/\sqrt{n}$, longer measurements and higher intensity can
increase precision. We may also increase precision with more interaction between probe and sample via multiple passes~\cite{Hi-Nat-450-393,birchall2017quantum} or optimising sample length~\cite{PhysRevResearch.2.033243}. However, there can often exist restrictions on the total optical exposure, the  measurement time and sample properties~\cite{cole2014live}. By using non-classical light, the fluctuations in the probe can be reduced below the QNL, thus providing `sub-shot-noise' precision per photon~\cite{xiao1987precision}. The QNL defines the best precision achievable without the use of quantum correlations for a given apparatus and average photon number~\cite{taylor2013biological}. This is distinguished from the standard quantum limit (SQL), which defines a measurement-independent limit to the precision that may be achieved using a minimum uncertainty state of a given average photon number, without quantum resources~\cite{gardiner2004quantum}. Because squeezed light can offer significant reduction in noise below the QNL~\cite{vahlbruch2016detection}, and can be generated with arbitrary intensity using coherent laser light~\cite{andersen201630},
it offers a practical approach for enhancing optical techniques beyond classical limitations. 

Precision in measuring a parameter 
can be quantified by the inverse of the variance of corresponding measurement outcomes, and is bounded by the Fisher information according to the Cramér-Rao bound~\cite{zwierz2010general}. By contrast, the sensitivity of a measurement is the smallest possible signal that may be observed~\cite{minkoff2002signal}, and thus depends only on the signal-to-noise ratio (SNR). 
Photonic (definite photon number) quantum metrology uses photon counting to observe
quantum correlations between modes to reduce the noise of a measurement~\cite{slussarenko2017unconditional,sabines2017sub,br-nphot-4-227}. Here one can attain a quantum advantage in both precision and sensitivity, by
increasing the Fisher information and  reducing the optical noise floor. However, due to limitations in both the maximum photon flux and detector saturation power,
the probe powers achievable 
are in practice
$\mathcal{O}(10^6)$ photons detected per second (pW)~\cite{slussarenko2017unconditional,sabines2017sub}, which limits use to only cases that are reliant on 
ultra low intensities.
Homodyne detection of squeezed vacuum has been used to estimate phase with sub-QNL precision~\cite{berni2015ab,yonezawa2012quantum}. This is possible because measurement is performed away from low frequency technical noise, in a shot-noise limited bandwidth where squeezing reduces vacuum noise.
However, as with photonic quantum metrology,
strategies using squeezed vacuum for sub-QNL
precision have also been restricted in 
maximum optical probe power.

\begin{figure*}[t]
\centering
\includegraphics[width=\textwidth]{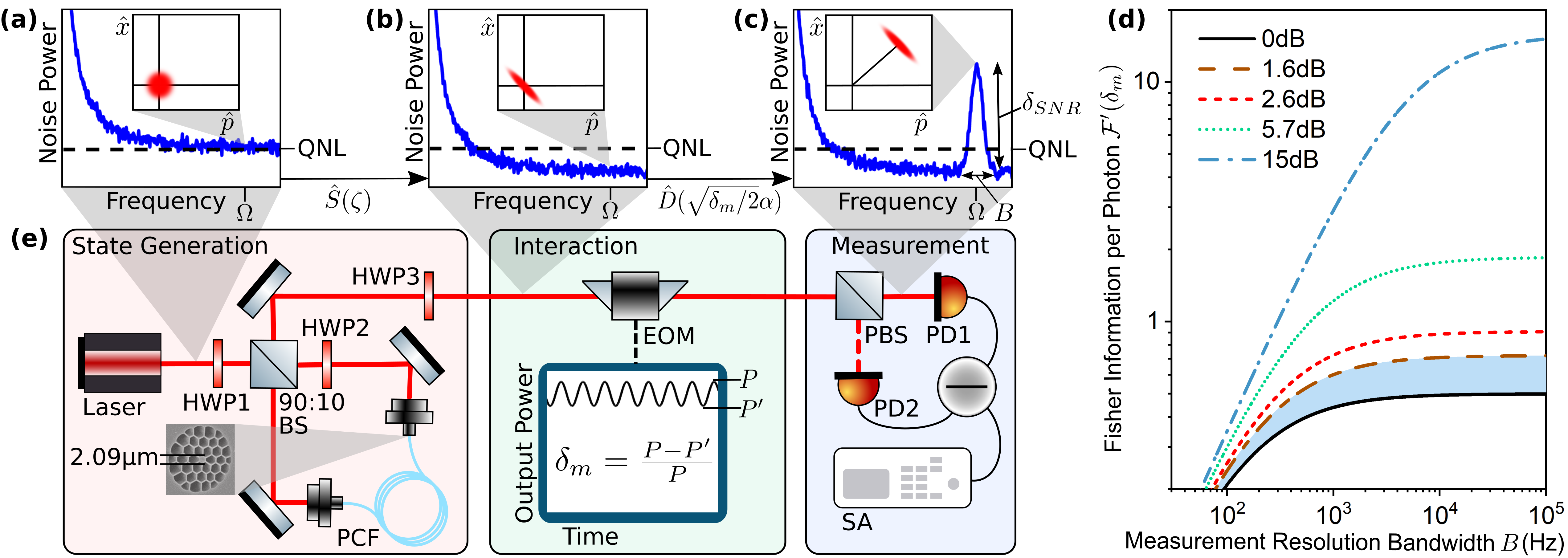}
\caption{Modelling enhanced precision measurement of $\delta_m$, and experimental setup. (a-c)  Plots of spectral noise power illustrating the effect of amplitude squeezing and modulation on a typical laser source, with the quadrature diagrams showing a coherent state defined by $\hat{x}$, $\hat{p}$ at $\pm\Omega$. (d) Theoretical model of the Fisher information per detected photon $\mathcal{F}'(\delta_m)$ for typical laser light which is quantum noise limited at $\Omega$ (solid line) and squeezed light (dashed lines): $-1.6$~dB and $-2.6$~dB are the measured and inferred generated squeezing levels in our experiment, $-5.7$~dB is amplitude squeezing previously achieved using an asymmetric Kerr interferometer~\cite{krylov1998amplitude} and $-15$~dB is the highest measured squeezing to date~\cite{vahlbruch2016detection}. For each plot, $M=1$, $P=0.2$~mW, $\lambda=740$~nm, $\eta=1$, $\delta_m=1\times10^{-4}$ and $\Var(\mathfrak{R}[\mathcal{H}])=1\times10^{-5}$. (e) Schematic of the experiment and SEM image representative of the PCF structure. A pulsed laser at $740$~nm propagates into the Sagnac interferometer for squeezed state generation. A birefringent photonic crystal fibre (PCF) provides the nonlinear medium for Kerr squeezing. The electro-optic modulator (EOM) combined with the polarising beamsplitter (PBS) are used to generate AM, which is measured on a spectrum analyser (SA).
}
\end{figure*}

Measurements using high power squeezed light can reach sub-QNL sensitivity in detection of phase modulation~\cite{xiao1987precision} and amplitude modulation (AM)~\cite{xiao1988detection}.
This is because modulation 
introduces AC components in the detected signal, which can be made to coincide with a shot-noise-limited detector bandwidth, while squeezed light reduces the optical noise relative to the signal. 
The sensitivity of any frequency domain measurement of an optical signal in a shot-noise limited bandwidth may be improved by such techniques and
this has been demonstrated in
a range of applications (e.g.~\cite{grangier1987squeezed,polzik1992spectroscopy,kasapi1997amplitude,li1997sub,sorensen1998quantum,taylor2013biological,wolfgramm2010squeezed,pooser2015ultrasensitive,PhysRevLett.123.231107,PhysRevLett.123.231108,de2020quantum,casacio2020quantum,garces2020quantum}). However, enhancing sensitivity is not a sufficient condition to enhance precision.
When bright optical probes are used,
the noise in the bright field often dominates over the vacuum noise, which prohibits the
use of squeezed light for reaching precision beyond the QNL.
For squeezed light to provide a precision improvement in such a measurement, the variance of the measured signal must be limited by optical shot-noise. 
Here, we fulfil this condition and use bright amplitude squeezed light to measure modulated transmission with precision beyond the QNL.

The parameter estimated in this work is the modulation index 
$\delta_m=(P-P')/P$, where $P$ 
and $P'$ are the maximum and minimum output power
transmitted through a modulated loss (Fig.~1(e))~\cite{bachor2004guide}. For 
modulation frequency $\Omega$, 
sinusoidal AM generates optical sidebands at $\pm\Omega$ from the carrier frequency. Upon photodetection, 
this leads to a single electronic sideband in the spectral noise power at frequency $\Omega$
that contains information about $\delta_m$. 
Figure~1(a-c) illustrate the behaviour of the spectral noise power of an initial laser input (a), where the noise characteristics at $\Omega$ approximate that of a coherent state $\ket{\alpha}$ and so quantum noise dominates the variance of intensity. The light is subsequently squeezed in amplitude (b) and then modulated in amplitude (c).  The insets illustrate the ideal evolution of the state at $\pm\Omega$ for an initial coherent state $\ket{\alpha}$. The final state is amplitude squeezed with an average photon number of $\langle \hat{n}(\pm\Omega)\rangle=\delta_m|\alpha|^2/2$.

We derive an estimator for $\delta_m$ from the SNR of direct photodetection, similar to~\cite{xiao1988detection} which uses
homodyne detection.
For direct photodetection of AM in a shot-noise limited bandwidth around $\Omega$, the 
SNR is given by $\delta_{SNR}=\langle p_s\rangle/\langle p_n\rangle$, where $\langle p_s\rangle$ is the average signal component of the generated electronic power at
$\Omega$, and $\langle p_n\rangle$ is the average electronic power generated from optical noise. 
In the limit of weak AM
($\delta_m\ll1$) loss due to AM has a negligible effect on 
the squeezing parameter $\Phi$ and the average optical power on the modulator output.
Therefore, average measured photocurrent 
is expressed as 
$i_0 = q\eta P(1-(\delta_m/2))/\hbar\omega \approx q\eta P/\hbar\omega$, with
electron charge $q$, photodiode efficiency $\eta$,  
reduced Planck constant $\hbar$ and 
carrier angular frequency $\omega$.
We then obtain
\begin{equation}\label{eq:SNR}
    \delta_{SNR} = \frac{\langle p_s\rangle}{\langle p_n\rangle} \approx \frac{\delta_m^2i_0}{4q\Phi B},
\end{equation}
(see Appendix Section 1), where $B$ is the frequency resolution bandwidth (RBW) of the noise spectrum,
and corresponds to the inverse of the integration time over which the spectrum is measured. From Eq.~(\ref{eq:SNR}), we 
define the estimator
\begin{equation}\label{eq:Estimator}
    \hat{\delta}_m = \sqrt{\frac{4q\Phi B\hat{\delta}_{SNR}}{i_0}},
\end{equation}
where
\begin{equation}\label{eq:EstimatorTerms}
          \hat{\delta}_{SNR}=\frac{p_\Omega-p_N}{p_N-p_E},
         \text{\hspace{0.3cm}and\hspace{0.3cm}}i_0 = \frac{q\eta\langle P\rangle}{\hbar\omega}.
\end{equation}
$p_\Omega$, $p_N$ and $p_E$ are the spectral noise powers of the electronic sideband, the optical noise floor and the electronic noise floor respectively.
$\langle P\rangle$ 
is the average optical power output from the modulator, and both $\langle P\rangle$ and $p_N$ may be
pre-calibrated with high precision. 
The dependence of $\hat{\delta}_m$ on the optical noise is then contained in the measurement of $p_\Omega$. 

For an input resistance of $R$ to the measuring device (e.g. spectrum analyser or oscilloscope), we can define the power of the electronic sideband as
\begin{equation}
    p_\Omega=2R|\hat{i}_\Omega|^2,
\end{equation}
where $\hat{i}_\Omega$ is the photocurrent in the frequency bin centered on $\Omega$. By considering power fluctuations due to quantum optical noise, low frequency classical optical noise, and electronic noise, we find
\begin{multline}\label{eq:VarPower}
    \Var(p_\Omega)=\langle p_\Omega^2\rangle-\langle p_\Omega\rangle^2\\\approx \frac{R^2}{M}\left[\vphantom{\frac{R^2}{M}}\right.2q\delta_m^2i_0^3\Phi B+ 4\delta_m^4i_0^4\text{Var}(\mathfrak{R}[\mathcal{H}])\\+4q^2\delta_m^2i_0^2\text{Var}(\mathfrak{R}[\mathcal{N}])\vphantom{R^2}\left.\vphantom{\frac{R^2}{M}}\right]
\end{multline}
(see Appendix Section 2). $\mathfrak{R}[\bullet]$ corresponds to the real part, $\mathcal{H}$ is the DC component of the classical relative amplitude noise from the laser, $\mathcal{N}$ is the component of electronic noise in the $\pm B/2$ frequency interval around $\Omega$, and $M$ is a variance reduction factor due to spectral averaging. The dependence of $\Var(p_\Omega)$ on $\mathcal{H}$ is due to classical noise being transferred from the carrier to the optical sidebands upon modulation. We assume here that the variance of the optical noise due to the classical intensity fluctuations scales quadratically with optical power, as expected for technical laser noise~\cite{svelto2010principles}. To quantify any advantage in precision obtained by using squeezed light, we compute the classical Fisher information on $\delta_m$, $\mathcal{F}(\delta_m)$. Since we use an amplitude squeezed state to perform an amplitude measurement, the classical Fisher information saturates the quantum Cramér-Rao bound~\cite{pinel2013quantum} --- therefore evaluating $\mathcal{F}(\delta_m)$ bounds any quantum strategy. For our measurement strategy, and assuming $\alpha\gg1$, $\hat{\delta}_{SNR}$ is normally distributed and we can define $\mathcal{F}(\delta_{SNR})$ according to~\cite{lehmann2006theory} 
\begin{equation}\label{eq:FisherInfoSNR}
    \mathcal{F}(\delta_{SNR})=\frac{1}{\Var(\hat{\delta}_{SNR})}=\left[\left(\pdv{\la\hat{\delta}_{SNR}\ra}{\la p_\Omega\ra}\right)^2\Var(p_\Omega)\right]^{-1}.
\end{equation}
$\mathcal{F}(\delta_m)$ can be obtained from $\mathcal{F}(\delta_{SNR})$ by using~\cite{lehmann2006theory}
\begin{equation}\label{eq:FisherInfoProp}
    \mathcal{F}(\delta_m)=\left(\pdv{\delta_{SNR}}{\delta_m}\right)^2\mathcal{F}(\delta_{SNR}).
\end{equation}
We find that $\Var(\mathfrak{R}[\mathcal{N}])$ contributes negligibly to $\mathcal{F}(\delta_m)$, and from Eq.~(\ref{eq:Estimator}-\ref{eq:FisherInfoProp}), this leads to
\begin{equation}\label{eq:FisherInfo}
    \mathcal{F}(\delta_m)\approx M\left[\frac{2q\Phi B}{i_0}+4\delta_m^2\Var(\mathfrak{R}[\mathcal{H}])\right]^{-1}
\end{equation}
The quantum advantage is then the ratio $\mathcal{Q}(\delta_m)$ between the values of $\mathcal{F}(\delta_m)$ for squeezed ($\Phi<1$) and coherent ($\Phi=1$) light.  The variance of $\hat{\delta}_m$ can be obtained by standard error propagation. We find
\begin{equation}\label{eq:ErrorProp}
    \text{Var}(\hat{\delta}_m)=\left(\frac{\partial \la\hat{\delta}_m\ra}{\partial \la p_\Omega\ra}\right)^2\text{Var}(p_\Omega)=\frac{1}{\mathcal{F}(\delta_m)}.
\end{equation}
Therefore, $\hat{\delta}_m$ is an efficient estimator. We also find that, in the limit of weak AM, $\la\hat{\delta}_m\ra=\delta_m$, meaning our estimator is unbiased.

The Fisher information per detected photon may be defined as $\mathcal{F}'(\delta_m)=\mathcal{F}(\delta_m)/\langle N\rangle$, where $\langle N\rangle=i_0/qB$ is the number of photons detected in the measurement time $B^{-1}$. Figure~1(d) illustrates the dependence of $\mathcal{F}'(\delta_m)$ on the RBW for a typical laser source which is quantum noise limited at $\Omega$ (solid black line) and various levels of squeezing (dashed lines), with all other parameters fixed. We find that for higher RBWs, squeezing provides sub-QNL precision in estimating $\delta_m$. This can be seen from Eq.~(\ref{eq:FisherInfo}), since for $2q\Phi B/i_0\gg4\delta_m^2\Var(\mathfrak{R}[\mathcal{H}])$, quantum noise limits the precision of the measurement, and we find $\mathcal{Q}(\delta_m)\rightarrow\mathcal{Q}_{opt}$, where
\begin{equation}\label{eq:QAopt}
    \mathcal{Q}_{opt}=\frac{1}{\Phi}.
\end{equation}
Because all information on $\delta_m$ is contained at modulation frequency $\Omega$, this model suggests a practically achievable quantum advantage per detected photon. 

\begin{figure*}[t]
\centering
\includegraphics[width=\textwidth]{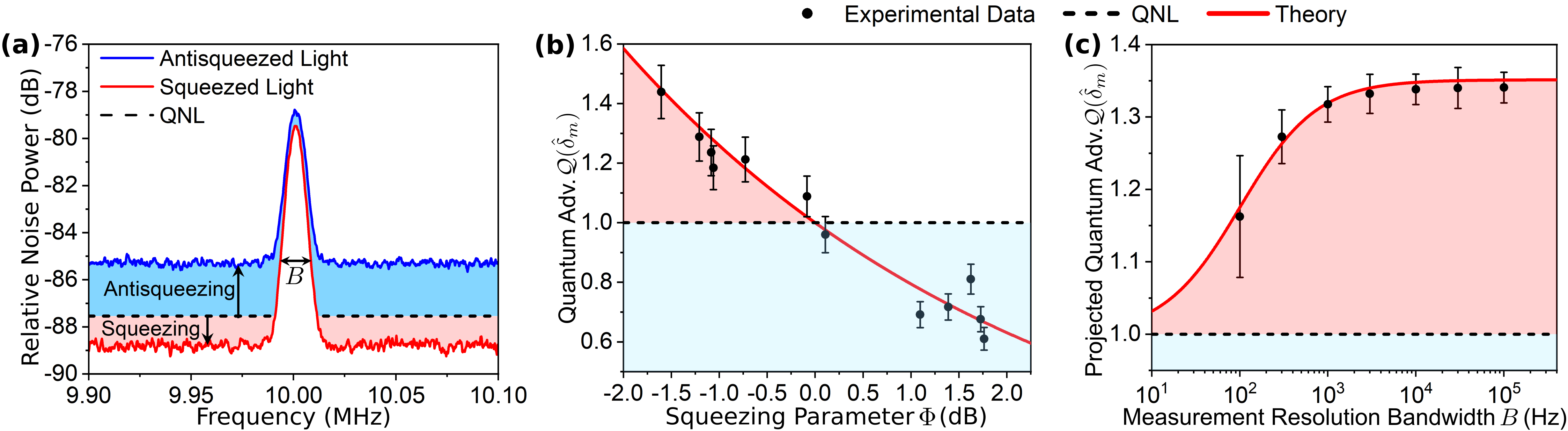}
\caption{Observing a quantum advantage in parameter estimation using amplitude squeezed light. In each plot, the black dashed line represents the QNL. (a) $10$~MHz AM measured by direct detection. The red trace corresponds to $-1.2$~dB of squeezing, and the blue trace to $2.7$~dB of antisqueezing. (b) Measured quantum advantage in precision of experimentally estimated $\delta_m$, $\mathcal{Q}(\hat{\delta}_m)$, for different squeezing $\Phi$. The red line corresponds to $\mathcal{Q}_{opt}$. (c) Measured $\mathcal{Q}(\hat{\delta}_m)$ with varied RBW $B$, for an average $-1.3$~dB of squeezing.
}
\end{figure*}

For the measurement, we built a source of amplitude squeezed light, based on~\cite{schmitt1998photon}.
$100$~fs pulses with central wavelength $\lambda_0=740$~nm from a Spectra Physics Mai Tai Ti:Sapphire laser 
are coupled into an asymmetric Sagnac interferometer, with a 90:10 splitting ratio beamsplitter (BS) (Fig.~1(e)). $14$~m of photonic crystal fibre (PCF) provides a strong $\chi^{(3)}$ nonlinearity in the interferometer. The fibre samples used were originally fabricated for photon pair generation work~\cite{alibart2006photon}. As the brighter 90\% reflected pulses propagate through the PCF, they undergo self-phase modulation and become quadrature squeezed~\cite{schmitt1998photon,andersen201630}. These pulses interfere with the weaker (10\%) counter-propagating pulses transmitted initially at the BS --- these provide a coherent displacement in phase space. This leads to amplitude squeezing on the output of the interferometer~\cite{schmitt1998photon}. The chosen central wavelength of $\lambda_0=740$~nm is close to the $730$~nm zero-dispersion wavelength of the PCF, in order to minimise the spectral broadening, which enables optimal interference at the 90:10 BS. The zero-dispersion wavelength of the PCF may be tailored by the fibre structure, making this approach applicable to a large range of wavelengths. The average optical power of the output state is $0.2$~mW, which equates to $25$~W of peak power. The amplitude squeezed light passes through a Thorlabs EO-AM-NR-C1 electro-optic modulator (EOM), that modulates polarisation.
A subsequent polarising beamsplitter (PBS) translates the polarisation modulation into a weak AM of depth $\delta_m$, and generates optical sidebands at distance $\pm\Omega$ from the carrier frequency. The resulting state is measured with direct detection, by collecting all the light at photodiode PD1. We calibrate the shot-noise level using the balanced subtraction photocurrent of PD1 and PD2. The balanced amplified photodetector used is a Thorlabs PDB440A(-AC). The spectral photocurrent is analysed with a Rohde \& Schwarz FPC1000 spectrum analyser.

Figure~2(a) shows the relative noise power traces of amplitude modulated squeezed light ($-1.2$~dB) and antisqueezed light ($2.7$~dB) produced by the setup.
The RBW is $B=10$~kHz, which is considerably wider than the linewidth of the optical sidebands, measured to be 
$<1$~Hz. The frequency separation of trace points in Fig.~2(a) is smaller than the RBW since the trace is a result of multiple samples within each RBW interval. This measurement demonstrates enhanced sensitivity detection of AM due to amplitude squeezing, as shown in~\cite{xiao1988detection}. The antisqueezed data in Fig.~2(a) is corrected for the difference in optical power required to generate antisqueezing and squeezing, by subtracting the difference in the respective shot-noise levels from this trace. The electronic noise has also been subtracted from each trace. 

From Eq.~(\ref{eq:ErrorProp}) we know that $\Var(\hat{\delta}_m)$ is proportional to $\Phi$ and inversely proportional to $\langle P\rangle$. However, the profile of squeezing with optical power is such that change in power is negligible across the maximum observed squeezing range $[-1.6,2.7]$~dB in our setup, 
so here $\Var(\hat{\delta}_m)$ scales linearly
with $\Phi$. By fitting measured $\Var(\hat{\delta}_m)$ to a line, we infer measured quantum advantage using
\begin{equation}\label{eq:QuantumAdv}
     \mathcal{Q}(\hat{\delta}_m)=\frac{\Var(\hat{\delta}_m)_{QNL}}{\Var(\hat{\delta}_m)_\Phi},
\end{equation}
where $\Var(\hat{\delta}_m)_{QNL}$ and $\Var(\hat{\delta}_m)_\Phi$ are variances of estimates of $\delta_m$, for coherent
and squeezed light respectively.

Figure~2(b) shows measured
$\mathcal{Q}(\hat{\delta}_m)$ for a range of squeezing $\Phi$ with fixed RBW $B=100$~kHz. 
The value of $\delta_m$ had a small experimental drift which varied between $\delta_m=[0.8, 1.0]\times10^{-4}$ over the duration of the measurements. Each sample of the spectral noise power corresponds to $M\approx34$, due to the video bandwidth filter, and every measurement of $\Var(\hat{\delta}_m)_\Phi$ is taken from 50 samples of $\hat{\delta}_m$. Due to the high RBW, quantum noise limits the variance of $\hat{\delta}_m$, and the measurement saturates the optimal quantum bound $\mathcal{Q}_{opt}$ given by Eq.~(\ref{eq:QAopt}) (red curve). A quantum advantage of $\mathcal{Q}(\hat{\delta}_m)=1.44\pm0.09$ is observed with $-1.6$~dB of squeezing, in agreement with $\mathcal{Q}_{opt}=1.45$. 

By repeating this for a range of $B$ and a fixed $-1.3$~dB of average squeezing, we 
plot in Fig.~2(c) the dependence of $\mathcal{Q}(\hat{\delta}_m)$ on RBW. 
The red curve is a theoretical fitting calculated using Eq.~(\ref{eq:FisherInfo}), with $\Var(\mathfrak{R}[\mathcal{H}])$ as a fitting parameter, which gives $\Var(\mathfrak{R}[\mathcal{H}])=7\pm1\times10^{-6}$. We observe sub-QNL precision down to $B=100$~Hz. The maximum quantum advantage observed here is $\mathcal{Q}(\hat{\delta}_m)=1.34\pm0.07$, which again closely agrees with the
the optimal $\mathcal{Q}_{opt}=1.35$ for the average squeezing parameter of $\Phi=0.74$. In Fig.~2(b) and Fig.~2(c), the error bars were derived from standard propagation of error calculations, with the data averaged over 236 variance measurements. Reducing optical loss to measure higher squeezing~\cite{vahlbruch2016detection} enables greater enhancement in precision.
Accounting for detection efficiency (we measure $\eta_d=0.84$) and coupling efficiency between the interferometer output and the detector 
($\eta_{opt}=0.81$), we infer our maximum measured squeezing value of $-1.6$~dB corresponds to $-2.6$~dB of generated amplitude squeezing.

We have demonstrated quantum enhanced precision parameter estimation with bright squeezed amplitude light. Our model shows the degree of 
precision enhancement is dictated by the amount of squeezing, the RBW and the classical noise on the generated sidebands. This exemplifies that for measurements of high power optical signals, sub-shot-noise sensitivity does not alone imply sub-shot-noise precision. We verify our model with experiment, reporting up to a $44\%$ quantum advantage in precision in the estimation of the modulation index, per detected photon. This general demonstration motivates applications in areas such as spectroscopy~\cite{Whittaker2017} and imaging~\cite{Sabines2019}, where precision may determine the performance of the measurement, which can be improved by using squeezed light. We use amplitude squeezed light of $0.2$~mW average optical power ($25$~W peak power) as a probe. This  power is comparable to the photon dose required to induce a photophobic response in living cells~\cite{berthold2008channelrhodopsin}, therefore indicating this technique's relevance to biological measurements.

Supporting data is available on request.

We thank J. Mueller, P. Mosley, A. Politi, J. Rarity, N. Samantaray, W. Wadsworth, S. Wollmann and W. Bowen for helpful discussions. This work was supported by QuantIC - The UK Quantum Technology Hub in Quantum Imaging (EPSRC EP/T00097X/1).  G.S.A. was supported by the Quantum Engineering Centre for Doctoral Training (EPSRC EP/L015730/1). E.J.A. acknowledges fellowship support from EPSRC Doctoral Prize Fellowship (EP/R513179/1). J.C.F.M. acknowledges support from an EPSRC Quantum Technology Fellowship (EP/M024385/1) and European Research Council starting grant ERC-2018-STG 803665. 

All authors contributed to the initial design and conception of the experiment. G.S.A, E.J.A, G.F and A.R.M contributed to the experimental setup. G.S.A performed the data collection. G.S.A, G.F and E.J.A contributed to the theoretical analysis. All authors discussed the results and contributed to the preparation of the manuscript.

\bibliographystyle{unsrtnat}

\newpage

\onecolumngrid
\section{Appendix: \\Quantum Enhanced Precision Estimation of Transmission with Bright Squeezed Light}

\subsection{1. Calculation of the Signal-to-Noise Ratio}

Here we derive the expected value of the electronic power $p_\OM$ in the $\pm B/2$ frequency interval centered on the modulation frequency $\OM$, generated by the current $\hat{i}(t)$. We may write $p_\OM$ as 
\begin{equation}\label{eq:power}
    p_\OM=2R|\hat{i}_\OM|^2,
\end{equation}
where $R$ is the input resistance, $\hat{i}_\OM=\int^\OM \hat{i}(\nu)d\nu$ is the photocurrent in the frequency bin centered on $\OM$, using the notation $\int^\OM\equiv\int_{\OM-B/2}^{\OM+B/2}$, and the factor of $2$ comes from the integration over positive and negative frequencies. It is important to note that this differs from the standard definition of the power in a frequency band, $p_\OM=2R\int^\OM|i(\nu)|^2 d\nu$~\cite{bachor2004guide}. The reason for the definition used in Eq.~(\ref{eq:power}) is that measuring devices such as spectrum analysers and oscilloscopes are fundamentally voltage detectors, and therefore the displayed power level is computed from the voltage measured in a given frequency bin. This means that the integration of the photocurrent density effectively occurs before taking the absolute square. While the average value $\la p_\OM\ra$ does not significantly differ between these definitions, $\text{Var}(p_\OM)$ does. To keep the model consistent with the measurements obtained by our spectrum analyser, we will use Eq.~(\ref{eq:power}) as a definition for $p_\OM$. 

The amplitude of the optical field before modulation is applied may be written as $\hat{A}_0(t)=[1 + \zeta(t)]\AL_0e^{i\theta} + \A(t)$, where $\theta$ is the phase of the classical field, $\A(t)$ describes the quantum amplitude fluctuations and $\zeta(t)$ is a stochastic noise function which corresponds to the low frequency classical noise of the laser. We have assumed a continuous-wave amplitude $\AL_0$ for simplicity. The amplitude of the detected light after modulation may then be written as
\begin{equation}
    \hat{A}(t)=(\PZ + \PM\cos(2\pi\OM t))([1 + \zeta(t)]\AL e^{i\theta} + \A(t)),
\end{equation}
where $\PZ$ and $\PM$ are related to the modulation index by $\PZ=1-\delta_m/2$ and $\PM=\delta_m/2$, and the detection efficiency $\eta$ is modelled as an additional loss before detection, such that $\AL=\sqrt{\eta}\AL_0$. By making the assumption of large amplitude $\AL\gg1$ and small modulation $\PM\ll1$, we can approximate $\hat{A}(t)$ as
\begin{equation}\label{eq:ClassicalField}
    \hat{A}(t)\approx(\PZ + \PM\cos(2\pi\OM t))[1 + \zeta(t)]\AL e^{i\theta} + \A(t)\equiv|\AL(t)|e^{i\theta} + \A(t),
\end{equation}
where the effect of amplitude modulation on the quantum noise term has been neglected. The current at time $t$ may then be written as
\begin{equation}
    \hat{i}(t)=q\left(\hat{A}(t)\D\hat{A}(t) + \n(t)\right)=q\left(|\AL(t)|^2 + \sqrt{2}|\AL(t)|\x(t) + \hat{a}(t)\D\hat{a}(t) + \n(t)\right),
\end{equation}
where we have defined the quadrature operator $\x(t)=\frac{1}{\sqrt{2}}[\hat{a}(t)e^{-i\theta}+\hat{a}(t)\D e^{i\theta}]$ and the electronic noise term $\n(t)$ corresponds to the number of electrons generated independently of the optical field. The component of the photocurrent at frequency $\V$ is then given by
\begin{equation}
    \hat{i}(\V)=q\left[I(\V) + \sqrt{2}\int\AL(\M)\x(\V-\M)d\M + \int\hat{a}(-\M)\D\hat{a}(\V-\M)d\M\right],
\end{equation}
where $\int\equiv\int_{-\infty}^{\infty}$, and we have defined the unitary Fourier transforms
\begin{equation}
    I(\V)=\int\left(|\AL(t)|^2 + \n(t)\right)e^{-2\pi i\V t}dt
\end{equation} and 
\begin{equation}
    \hat{x}_\theta(\V)=\int \x(t)e^{-2\pi i\V t}dt=\frac{1}{\sqrt{2}}\left[\hat{a}(-\V)\D e^{i\theta} + \hat{a}(\V)e^{-i\theta}\right].
\end{equation}
We also define $\A(\V)$ as the squeezed vacuum operator~\cite{loudon2000quantum}
\begin{equation}
     \A(\V)=\hat{b}(\V)\cosh{r(\V)}-e^{2i\theta(\V)}\hat{b}(-\V)\D\sinh{r(\V)},
\end{equation}
where $\hat{b}(\V)$ and $\hat{b}(\V)\D$ are the bosonic creation and annihilation operators. The squeezing is defined such that $r(\V)=r$ and $\theta(\V)=\theta$ within the frequency bandwidth $-\Lambda/2\leq\V\leq\Lambda/2$ (where $\Lambda/2>\OM$) and $r(\V)=0$ outside of this frequency range. The $2\theta$ phase factor then orients the squeezing in the amplitude direction. By defining $I_\OM=\int^\OM I(\V)d\V$ we can write $p_\OM$ as
\begin{multline}
    p_\OM=2q^2R\left[\vphantom{\int}\right.|I_\OM|^2 + \sqrt{2}I_\OM^*\int^\OM\int\AL(\M)\x(\V-\M) d\M d\V + I_\OM^*\int^\OM\int\hat{a}(-\M)\D\hat{a}(\V-\M)d\M d\V\\ \sqrt{2}I_\OM\int^\OM\int\AL(\M)^*\x(\V-\M)\D d\M d\V +     2\int^\OM\int^\OM\int\int\AL(\M)^*\AL(\MB)\x(\V-\M)\D\x(\VB-\MB) d\M d\MB d\V d\VB\\ + \sqrt{2}\int^\OM\int^\OM\int\int\AL(\M)^*\x(\V-\M)\D\hat{a}(-\MB)\D\hat{a}(\VB-\MB) d\M d\MB d\V d\VB + I_\OM\int^\OM\int\hat{a}(\V-\M)\D\hat{a}(-\M)d\M d\V \\ + \sqrt{2}\int^\OM\int^\OM\int\int\AL(\M)\hat{a}(\V-\MB)\D\hat{a}(-\MB)\x(\VB-\M) d\M d\MB d\V d\VB \\+ \int^\OM\int^\OM\int\int\hat{a}(\V-\M)\D\hat{a}(-\M)\hat{a}(-\MB)\D\hat{a}(\VB-\MB)d\M d\MB d\V d\VB\left.\vphantom{\int}\right],
\end{multline}
where $(\bullet)^*$ denotes the complex conjugate. From Eq.~(\ref{eq:ClassicalField}), we can find the frequency dependence of the classical field amplitude:
\begin{equation}\label{eq:alpha}
    \AL(\V)=\int|\AL(t)|e^{-2\pi i\V t}=\AL\left[\PZ(\delta(\V) + h(\V)) +  \frac{\PM}{2}\left(\delta(\V-\OM) + \delta(\V+\OM) + h(\V-\OM) + h(\V+\OM)\right) \right],
\end{equation}
where $h(\V)=\int\zeta(t)e^{-2\pi i\V t}dt$, and since classical noise is only observed at low frequencies ($\lesssim2$~MHz), we can write for example $h(\OM)=0$. Equation~(\ref{eq:ClassicalField}) allows us to define $I(\V)$ as
\begin{multline}
    I(\V)=\AL^2\left[\vphantom{\int}\right.\PZ^2\left(\delta(\V) + 2h(\V) + \int h(\M)h(\V-\M)d\M\right) + \PZ\PM\left(\vphantom{\int}\right.\delta(\OM-\V) + \delta(\OM+\V) \\+ 2h(\V-\OM) + 2h(\V + \OM) + \int h(\M)h(\V-\M-\OM)d\M + \int h(\M)h(\V-\M+\OM)d\M\left.\vphantom{\int}\right) +
    \\+ \frac{\PM^2}{4}\left(\vphantom{\int}\right.\delta(\V-2\OM) + \delta(\V+2\OM) + 2\delta(\V) + 2h(\V-2\OM) + 2h(\V+2\OM) + 4h(\V) \\+ \int h(\M)h(\V-\M-2\OM)d\M + \int h(\M)h(\V-\M+2\OM)d\M + 2\int h(\M)h(\V-\M)d\M\left.\vphantom{\int}\right)\left.\vphantom{\int}\right]+\n(\V).
\end{multline}
We can then find the expectation $\la p_\OM\ra$ with respect to the random variables $h(\V)$, $\n(\V)$ and $\A(\V)$. Since these variables are independent, the expectation value may be defined as $\la\bullet\ra\equiv\la\la\bra{0}\bullet\ket{0}\ra_{h(\V)}\ra_{\n(\V)}$. To calculate this, we first compute
\begin{multline}\label{eq:int_I_squared}
    |I_\OM|^2=\PZ^2\PM^2\AL^4\left[\vphantom{\left|\int\right|^2}\right.1 + 4\int^\OM\R[h(\V-\OM)]d\V + 2\int^\OM\int\R[h(\M)h(\V-\M-\OM)]d\M d\V + 4\left|\int^\OM h(\V-\OM)d\V\right|^2 \\+ 4\int^\OM\int^\OM\int\R[h(\V-\OM)^*h(\M)h(\VB-\M-\OM)]d\M d\V d\VB + \left|\int^\OM\int h(\M)h(\V-\M-\OM)d\M d\V\right|^2\left.\vphantom{\left|\int\right|^2}\right] \\+ \PZ\PM\AL^2\left[\vphantom{\left|\int\right|^2}\right. 2\int^\OM\R[\n(\V)]d\V + 4\int^\OM\int^\OM \R[h(\V-\OM)^*\n(\VB)]d\V d\VB \\+ 2\int^\OM\int^\OM\int\R[h(\M)^*h(\V-\M-\OM)^*\n(\VB)]d\M d\V d\VB \left.\vphantom{\left|\int\right|^2}\right] + \left|\int^\OM\n(\V)d\V\right|^2,
\end{multline}
where $\R[\bullet]$ signifies the real part. Then, by evaluating the quantum part of the expectation value, we obtain the result:
\begin{multline}\label{eq:av_power}
    \la p_\OM\ra=2q^2R\left[\la|I_\OM|^2\ra +\Phi\int^\OM\int^\OM\int\la\AL(\M)^*\AL(\M+\V-\VB)\ra d\M d\V d\VB+B\Lambda\left(\frac{\Phi^2}{8}+\frac{1}{8\Phi^2}-\frac{1}{4}\right)\right] \\ = 2q^2R\left[\vphantom{\left\la\left|\int\right|^2\right\ra}\right.\AL^4\left(\vphantom{\left\la\left|\int\right|^2\right\ra}\right.\PZ^2\PM^2 + 4\PZ^2\PM^2\int^\OM\la\R[h(\V-\OM)]\ra d\V + 2\PZ^2\PM^2\int^\OM\int\la\R[h(\M)h(\V-\M-\OM)]\ra d\M d\V \\+ 4\PZ^2\PM^2\left\la\left|\int^\OM h(\V-\OM) d\V\right|^2\right\ra + 4\PZ^2\PM^2\int^\OM\int^\OM\int\la\R[h(\V-\OM)^*h(\M)h(\VB-\M-\OM)]\ra d\M d\V d\VB \\+ \PZ^2\PM^2\left\la\left|\int^\OM\int h(\M)h(\V-\M-\OM) d\M d\V\right|^2\right\ra\left.\vphantom{\left\la\left|\int\right|^2\right\ra}\right) + \AL^2\left(\vphantom{\left\la\left|\int\right|^2\right\ra}\right.2\PZ\PM\int^\OM\la\R[\n(\V)]\ra d\V\\ +4\PZ\PM\int^\OM\int^\OM \la\R[h(\V-\OM)^*\n(\VB)]\ra d\V d\VB + 2\PZ\PM\int^\OM\int^\OM\int\left\la\R[h(\M)^*h(\V-\M-\OM)^*\n(\VB)]\right\ra d\M d\V d\VB\\ + \left(\PZ^2+\frac{\PM^2}{2}\right)\Phi B  + (2\PZ^2+\PM^2)\Phi\int^\OM\int^\OM\la\R[h(\V-\VB)]\ra d\V d\VB\\ + \left(\PZ^2+\frac{\PM^2}{2}\right)\Phi\int^\OM\int^\OM\int\la h(\M)^*h(\M+\V-\VB)\ra d\M d\V d \VB \left.\vphantom{\left\la\left|\int\right|^2\right\ra}\right)+ \left\la\left|\int^\OM\n(\V)d\V\right|^2\right\ra \\+ B\Lambda\left(\frac{\Phi^2}{8}+\frac{1}{8\Phi^2}-\frac{1}{4}\right)\left.\vphantom{\left\la\left|\int\right|^2\right\ra}\right],
\end{multline}
for the squeezing parameter $\Phi=e^{-2r}$. In Eq.~(\ref{eq:av_power}), we have neglected terms involving the expectation value of the product of an odd number of creation or annihilation operators, and terms outside the domain of $h(\V)$. By observing that $\AL\gg1$, $\la|\int h(\V) d\V|\ra\ll1$ and $\delta_m\ll1$ for $i_0\approx q\AL^2$ we find
\begin{equation}
    \la p_\OM\ra\approx R\left(\frac{i_0^2\delta_m^2}{2}+2qi_0\Phi B+2q^2\left\la\left|\int^\OM\n(\V)d\V\right|^2\right\ra\right).
\end{equation}
Similarly, at a small frequency interval $\Delta f$ from $\OM$, we find that the electronic power of the optical noise floor and electronic noise floor are respectively
\begin{equation}
    \la p_N\ra\approx R\left(2qi_0\Phi B+2q^2\left\la\left|\int^\OM\n(\V)d\V\right|^2\right\ra\right) \qquad\text{and}\qquad \la p_E\ra=R\left(2q^2\left\la\left|\int^\OM\n(\V)d\V\right|^2\right\ra\right).
\end{equation}
We then find that the optical signal-to-noise ratio $\delta_{SNR}$ may be expressed as
\begin{equation}
    \delta_{SNR}=\frac{ \la p_\OM\ra-\la p_N\ra}{\la p_N\ra-\la p_E\ra}\approx\frac{i_0\delta_m^2}{4q\Phi B}.
\end{equation}

\subsection{2. Calculation of the Variance of the Sideband Power}

In order to calculate the variance $\text{Var}(p_\OM)=\la p_\OM^2\ra-\la p_\OM\ra^2$, an expression for $\la p_\OM\ra^2$ may be evaluated directly from Eq.~(\ref{eq:av_power}) to give
\begin{multline}\label{eq:av_p_squared}
    \la p_\OM\ra^2=4q^4R^2\Bigg[\la|I_\OM|^2\ra^2 + (2\PZ^2+\PM^2)\la|I_\OM|^2\ra\AL^2\left(\vphantom{\int}\right.\Phi B + 2\Phi\int^\OM\int^\OM\la\R[h(\V-\VB)]\ra d\V d\VB \\ + \Phi\int^\OM\int^\OM\int \la h(\M)^*h(\M+\V-\VB)\ra d\M d\V d\VB \left.\vphantom{\int}\right) + \la|I_\OM|^2\ra\Lambda B\left(\frac{\Phi^2}{4}+\frac{1}{4\Phi^2}-\frac{1}{2}\right) \\+ \left(\PZ^4+\PZ^2\PM^2+\frac{\PM^4}{4}\right)\AL^4\left(\vphantom{\int}\right.\Phi^2 B^2 + 4\Phi^2B\int^\OM\int^\OM\la\R[h(\V-\VB)]\ra d\V d\VB \\+ 2\Phi^2B\int^\OM\int^\OM\int\la h(\M)^*h(\M+\V-\VB)\ra d\M d\V d\VB + 4\Phi^2\int^\OM\int^\OM\int^\OM\int^\OM\la\R[h(\V-\VB)]\ra \la\R[h(\VBB-\VBBB)]\ra d\V d\VB d\VBB d\VBBB \\ + 4\Phi^2\int^\OM\int^\OM\int^\OM\int^\OM\int\la\R[h(\V-\VB)]\ra\la h(\M)^*h(\M+\VBB-\VBBB)\ra d\M d\V d\VB d\VBB d\VBBB \\ +  \Phi^2\int^\OM\int^\OM\int^\OM\int^\OM\int\int\la h(\M)^*h(\M+\V-\VB)\ra\la h(\MB)^*h(\MB+\VBB-\VBBB)\ra d\M d\MB d\V d\VB d\VBB d\VBBB \left.\vphantom{\int}\right)\\+ (2\PZ^2+\PM^2)\AL^2\left(\frac{\Phi^2}{8}+\frac{1}{8\Phi^{2}}-\frac{1}{4}\right)\left(\vphantom{\int}\right.\Phi \Lambda B^2 + 2\Phi\Lambda B\int^\OM\int^\OM\la\R[h(\V-\VB)]\ra d\V d\VB \\ + \Phi\Lambda B\int^\OM\int^\OM\int \la h(\M)^*h(\M+\V-\VB)\ra d\M d\V d\VB  \left.\vphantom{\int}\right) + \Lambda^2 B^2\left(\frac{\Phi^2}{8}+\frac{1}{8\Phi^{2}}-\frac{1}{4}\right)^2 \Bigg].
\end{multline}
To find an expression for $\la p_\OM^2\ra$, we can again neglect terms where the expectation value of the quadrature operators vanishes, giving 
\begin{multline}\label{eq:exp(p_squared)}
    \la p_\OM^2\ra=4q^4R^2\left\la\vphantom{\int}\right.|I_\OM|^4 + 4|I_\OM|^2\int^\OM\int^\OM\int\int\AL(\M)^*\AL(\MB)\x(\V-\M)\D\x(\VB-\MB)d\M d\MB d\V d\VB \\ +  2I_\OM^{*2}\int^\OM\int^\OM\int\int\AL(\M)\AL(\MB)\x(\V-\M)\x(\VB-\MB)d\M d\MB d\V d\VB\\ + 2I_\OM^2\int^\OM\int^\OM\int\int\AL(\M)^*\AL(\MB)^*\x(\V-\M)\D\x(\VB-\MB)\D d\M d\MB d\V d\VB\\ + 4|I_\OM|^2\int^\OM\int^\OM\int\int\AL(\M)\AL(\MB)^*\x(\V-\M)\x(\VB-\MB)\D d\M d\MB d\V d\VB\\ +  3|I_\OM|^2\int^\OM\int^\OM\int\int\A(\V-\M)\D\A(-\M)\A(-\MB)\D\A(\VB-\MB)d\M d\MB d\VB d\VBB \\ + |I_\OM|^2\int^\OM\int^\OM\int\int\A(-\M)\D\A(\V-\M)\A(\VB-\MB)\D\A(-\MB)d\M d\MB d\V d\VBB\\ + 2I_\OM^*\int^\OM\int^\OM\int^\OM\int\int\int \AL(\M)\AL(\MB)^*\x(\V-\M)\x(\VB-\MB)\D\A(-\MBB)\D\A(\VBB-\MBB)d\M d\MB d\MBB d\V d\VB d\VBB\\ + 2I_\OM^*\int^\OM\int^\OM\int^\OM\int\int\int \AL(\M)\AL(\MB)^*\x(\VB-\MB)\D\A(-\MBB)\D\A(\VBB-\MBB)\x(\V-\M)d\M d\MB d\MBB d\V d\VB d\VBB\\ + 2I_\OM^*\int^\OM\int^\OM\int^\OM\int\int\int \AL(\M)^*\AL(\MB)\A(-\MBB)\D\A(\V-\MBB)\x(\VB-\M)\D\x(\VBB-\MB)d\M d\MB d\MBB d\V d\VB d\VBB\\ + 2I_\OM^*\int^\OM\int^\OM\int^\OM\int\int\int \AL(\M)^*\AL(\MB)\x(\VB-\M)\D\x(\VBB-\MB)\A(-\MBB)\D\A(\V-\MBB)d\M d\MB d\MBB d\V d\VB d\VBB\\ + 2I_\OM\int^\OM\int^\OM\int^\OM\int\int\int \AL(\M)^*\AL(\MB)\x(\V-\M)\D\A(\VB-\MBB)\D\A(-\MBB)\x(\VBB-\MB)d\M d\MB d\MBB d\V d\VB d\VBB\\ + 2I_\OM\int^\OM\int^\OM\int^\OM\int\int\int \AL(\M)^*\AL(\MB)\A(\VB-\MBB)\D\A(-\MBB)\x(\VBB-\MB)\x(\V-\M)\D d\M d\MB d\MBB d\V d\VB d\VBB\\ + 2I_\OM\int^\OM\int^\OM\int^\OM\int\int\int \AL(\M)^*\AL(\MB)\x(\V-\M)\D\x(\VB-\MB)\A(\VBB-\MBB)\D\A(-\MBB)d\M d\MB d\MBB d\V d\VB d\VBB\\ + 2I_\OM\int^\OM\int^\OM\int^\OM\int\int\int \AL(\M)^*\AL(\MB)\A(\VBB-\MBB)\D\A(-\MBB)\x(\V-\M)\D\x(\VB-\MB)d\M d\MB d\MBB d\V d\VB d\VBB\\ + 2I_\OM^*\int^\OM\int^\OM\int^\OM\int\int\int \AL(\M)\AL(\MB)\x(\V-\M)\A(\VB-\MBB)\D\A(-\MBB)\x(\VBB-\MB)d\M d\MB d\MBB d\V d\VB d\VBB\\ + 2I_\OM^*\int^\OM\int^\OM\int^\OM\int\int\int \AL(\M)\AL(\MB)\A(\VB-\MBB)\D\A(-\MBB)\x(\VBB-\MB)\x(\V-\M)d\M d\MB d\MBB d\V d\VB d\VBB\\ + 2I_\OM\int^\OM\int^\OM\int^\OM\int\int\int \AL(\M)^*\AL(\MB)^*\x(\V-\M)\D\x(\VB-\MB)\D\A(-\MBB)\D\A(\VBB-\MBB)d\M d\MB d\MBB d\V d\VB d\VBB\\ + 2I_\OM\int^\OM\int^\OM\int^\OM\int\int\int \AL(\M)^*\AL(\MB)^*\x(\VB-\MB)\D\A(-\MBB)\D\A(\VBB-\MBB)\x(\V-\M)\D d\M d\MB d\MBB d\V d\VB d\VBB\\ + 4\int^\OM\int^\OM\int^\OM\int^\OM\int\int\int\int\AL(\M)^*\AL(\MB)\AL(\MBB)^*\AL(\MBBB)\x(\V-\M)\D\x(\VB-\MB)\x(\VBB-\MBB)\D\x(\VBBB-\MBBB)d\M d\MB d\MBB d\MBBB d\V d\VB d\VBB d\VBBB \\ + 2\int^\OM\int^\OM\int^\OM\int^\OM\int\int\int\int\AL(\M)^*\AL(\MB)\x(\V-\M)\D\x(\VB-\MB)\hat{a}(\VBB-\MBB)\D\hat{a}(-\MBB)\hat{a}(-\MBBB)\D\hat{a}(\VBBB-\MBBB)d\M d\MB d\MBB d\MBBB d\V d\VB d\VBB d\VBBB\\ + 2\int^\OM\int^\OM\int^\OM\int^\OM\int\int\int\int\AL(\M)^*\AL(\MB)\hat{a}(\VBB-\MBB)\D\hat{a}(-\MBB)\hat{a}(-\MBBB)\D\hat{a}(\VBBB-\MBBB)\x(\V-\M)\D\x(\VB-\MB)d\M d\MB d\MBB d\MBBB d\V d\VB d\VBB d\VBBB\\ + 2\int^\OM\int^\OM\int^\OM\int^\OM\int\int\int\int\AL(\M)^*\AL(\MB)\x(\V-\M)\D\hat{a}(-\MBB)\D\hat{a}(\VB-\MBB)\hat{a}(\VBB-\MBBB)\D\hat{a}(-\MBBB)\x(\VBBB-\MB)d\M d\MB d\MBB d\MBBB d\V d\VB d\VBB d\VBBB\\ + 2\int^\OM\int^\OM\int^\OM\int^\OM\int\int\int\int\AL(\M)^*\AL(\MB)\hat{a}(\VBB-\MBBB)\D\hat{a}(-\MBBB)\x(\VBBB-\MB)\x(\V-\M)\D\hat{a}(-\MBB)\D\hat{a}(\VB-\MBB)d\M d\MB d\MBB d\MBBB d\V d\VB d\VBB d\VBBB\\ + 2\int^\OM\int^\OM\int^\OM\int^\OM\int\int\int\int\AL(\M)^*\AL(\MB)^*\x(\V-\M)\D\hat{a}(-\MBB)\hat{a}(\VB-\MBB)\x(\VBB-\MB)\D\hat{a}(-\MBBB)\D\hat{a}(\VBBB-\MBBB)d\M d\MB d\MBB d\MBBB d\V d\VB d\VBB d\VBBB\\+ 2\int^\OM\int^\OM\int^\OM\int^\OM\int\int\int\int\AL(\M)\AL(\MB)\hat{a}(\V-\MBB)\D\hat{a}(-\MBB)\x(\VB-\M)\hat{a}(\VBB-\MBBB)\D\hat{a}(-\MBBB)\x(\VBBB-\MB)d\M d\MB d\MBB d\MBBB d\V d\VB d\VBB d\VBBB\\ + \int^\OM\int^\OM\int^\OM\int^\OM\int\int\int\int\A(\V-\M)\D\A(-\M)\A(-\MB)\D\A(\VB-\MB)\A(\VBB-\MBB)\D\A(-\MBB)\A(-\MBBB)\D\A(\VBBB-\MBBB)d\M d\MB d\MBB d\MBBB d\V d\VB d\VBB d\VBBB \left.\vphantom{\int}\right\ra.
\end{multline}
To explicitly evaluate $\la p_\OM^2\ra$ in the following, we calculate the integrals in Eq.~(\ref{eq:exp(p_squared)}) separately, using Eq.~(\ref{eq:alpha}) and the commutation relations of the bose operators with the expectation value taken on the vacuum state. Terms 2-5 of Eq.~(\ref{eq:exp(p_squared)}) respectively give:
\begin{multline}
    \left\la4|I_\OM|^2\int^\OM\int^\OM\int\int\AL(\M)^*\AL(\MB)\x(\V-\M)\D\x(\VB-\MB)d\M d\MB d\V d\VB\right\ra \\= (2\PZ^2+\PM^2)\AL^2\Phi\left(\vphantom{\int}\right.\la|I_\OM|^2\ra B + 2\int^\OM\int^\OM\la|I_\OM|^2\R[h(\V-\VB)]\ra d\V d\VB \\ + \int^\OM\int^\OM\int\la|I_\OM|^2h(\M)^*h(\M+\V-\VB)\ra d\M d\V d\VB\left.\vphantom{\int}\right),
\end{multline}

\begin{multline}
    \left\la2I_\OM^{*2}\int^\OM\int^\OM\int\int\AL(\M)\AL(\MB)\x(\V-\M)\x(\VB-\MB)d\M d\MB d\V d\VB\right\ra\\=\frac{\PM^2}{4}\AL^2\Phi\left(\vphantom{\int}\right.\la I_\OM^{*2}\ra B + 2\int^\OM\int^\OM\la I_\OM^{*2}h(\V+\VB-2\OM)\ra d\V d\VB \\+ \int^\OM\int^\OM\int\la I_\OM^{*2}h(\M-\OM)h(\V+\VB-\M-\OM)\ra d\M d\V d\VB\left.\vphantom{\int}\right),
\end{multline}

\begin{multline}
    \left\la2I_\OM^2\int^\OM\int^\OM\int\int\AL(\M)^*\AL(\MB)^*\x(\V-\M)\D\x(\VB-\MB)\D d\M d\MB d\V d\VB\right\ra\\=\frac{\PM^2}{4}\AL^2\Phi\left(\vphantom{\int}\right.\la I_\OM^2\ra B + 2\int^\OM\int^\OM\la I_\OM^2h(\V+\VB-2\OM)^*\ra d\V d\VB \\+ \int^\OM\int^\OM\int\la I_\OM^2h(\M-\OM)^*h(\V+\VB-\M-\OM)^*\ra d\M d\V d\VB\left.\vphantom{\int}\right)
\end{multline}
and
\begin{multline}
    \left\la4|I_\OM|^2\int^\OM\int^\OM\int\int\AL(\M)\AL(\MB)^*\x(\V-\M)\x(\VB-\MB)\D d\M d\MB d\V d\VB\right\ra \\= (2\PZ^2+\PM^2)\AL^2\Phi\left(\vphantom{\int}\right.\la|I_\OM|^2\ra B + 2\int^\OM\int^\OM\la|I_\OM|^2\R[h(\V-\VB)]\ra d\V d\VB \\ + \int^\OM\int^\OM\int\la|I_\OM|^2h(\M)^*h(\M+\V-\VB)\ra d\M d\V d\VB\left.\vphantom{\int}\right).
\end{multline}

The summation of terms 6 and 7 of Eq.~(\ref{eq:exp(p_squared)}) gives

\begin{multline}
    \left\la3|I_\OM|^2\int^\OM\int^\OM\int\int\A(\V-\M)\D\A(-\M)\A(-\MB)\D\A(\VB-\MB)d\M d\MB d\VB d\VBB\right\ra \\ + \left\la|I_\OM|^2\int^\OM\int^\OM\int\int\A(-\M)\D\A(\V-\M)\A(\VB-\MB)\D\A(-\MB)d\M d\MB d\V d\VBB\right\ra = \la|I_\OM|^2\ra\Lambda B\left(\frac{\Phi^2}{2}+\frac{1}{2\Phi^{2}}-1\right).
\end{multline}

It is also possible to combine terms 8-15 of Eq.~(\ref{eq:exp(p_squared)}) as follows:

\begin{multline}
    2\int^\OM\int^\OM\int^\OM\int\int\int\left[\vphantom{\int}\right.\la I_\OM^*\AL(\M)\AL(\MB)^*\x(\V-\M)\x(\VB-\MB)\D\A(-\MBB)\D\A(\VBB-\MBB)\ra\\ + 
    \la I_\OM^*\AL(\M)\AL(\MB)^*\x(\VB-\MB)\D\A(-\MBB)\D\A(\VBB-\MBB)\x(\V-\M)\ra\\ + 
    \la I_\OM^*\AL(\M)^*\AL(\MB)\A(-\MBB)\D\A(\V-\MBB)\x(\VB-\M)\D\x(\VBB-\MB)\ra\\ + 
    \la I_\OM^*\AL(\M)^*\AL(\MB)\x(\VB-\M)\D\x(\VBB-\MB)\A(-\MBB)\D\A(\V-\MBB)\ra\\ + 
    \la I_\OM\AL(\M)^*\AL(\MB)\x(\V-\M)\D\A(\VB-\MBB)\D\A(-\MBB)\x(\VBB-\MB)\ra\\ + 
    \la I_\OM\AL(\M)^*\AL(\MB)\A(\VB-\MBB)\D\A(-\MBB)\x(\VBB-\MB)\x(\V-\M)\D\ra\\ + 
    \la I_\OM\AL(\M)^*\AL(\MB)\x(\V-\M)\D\x(\VB-\MB)\A(\VBB-\MBB)\D\A(-\MBB)\ra\\ + 
    \la I_\OM\AL(\M)^*\AL(\MB)\A(\VBB-\MBB)\D\A(-\MBB)\x(\V-\M)\D\x(\VB-\MB)\ra\left.\vphantom{\int}\right]d\M d\MB d\MBB d\V d\VB d\VBB\\=(4\Phi^2-2)\int^\OM\int^\OM\int^\OM\int\la\R[I_\OM]\AL(\M)\AL(\M+\V-\VB-\VBB)^*\ra d\M d\VB d\VBB d\VBBB\\=\PZ\PM\AL^2(4\Phi^2-2)\left(\vphantom{\int}\right.\la\R[I_\OM]\ra B^2 + 2\int^\OM\int^\OM\int^\OM\la\R[I_\OM]\R[h(\V+\VB-\VBB-\OM)]\ra d\V d\VB d\VBB \\+ \int^\OM\int^\OM\int^\OM\int\la\R[I_\OM]h(\M-\OM)h(\M+\V-\VB-\VBB)^*\ra d\M d\V d\VB d\VBB \left.\vphantom{\int}\right).
\end{multline}

Similarly, we find that terms 16-19 of Eq.~(\ref{eq:exp(p_squared)}) simplify as
\begin{multline}
    2\int^\OM\int^\OM\int^\OM\int\int\int\left[\vphantom{\int}\right.\la I_\OM^*\AL(\M)\AL(\MB)\x(\V-\M)\A(\VB-\MBB)\D\A(-\MBB)\x(\VBB-\MB)\ra\\+\la I_\OM^*\AL(\M)\AL(\MB)\A(\VB-\MBB)\D\A(-\MBB)\x(\VBB-\MB)\x(\V-\M)\ra\\+
    \la I_\OM\AL(\M)^*\AL(\MB)^*\x(\V-\M)\D\x(\VB-\MB)\D\A(-\MBB)\D\A(\VBB-\MBB)\ra\\+
    \la I_\OM\AL(\M)^*\AL(\MB)^*\x(\VB-\MB)\D\A(-\MBB)\D\A(\VBB-\MBB)\x(\V-\M)\D\ra\left.\vphantom{\int}\right]d\M d\MB d\MBB d\V d\VB d\VBB\\=\Phi^2\int^\OM\int^\OM\int^\OM\int\la I_\OM^*\AL(\M)\AL(\M+\V-\VB-\VBB)\ra d\M d\VB d\VBB d\VBBB\\+\Phi^2\int^\OM\int^\OM\int^\OM\int\la I_\OM\AL(\M)^*\AL(\M+\V-\VB-\VBB)^*\ra d\M d\VB d\VBB d\VBBB\\=2\PZ\PM\AL^2\Phi^2\left(\vphantom{\int}\right.\la\R[I_\OM]\ra B^2 + 2\int^\OM\int^\OM\int^\OM\la\R[I_\OM^* h(\V+\VB-\VBB-\OM)]\ra d\V d\VB d\VBB \\+ \int^\OM\int^\OM\int^\OM\int\la\R[I_\OM^*h(\M-\OM)h(\M+\V-\VB-\VBB)]\ra d\M d\V d\VB d\VBB \left.\vphantom{\int}\right).
\end{multline}

We can write term 20 of Eq.~(\ref{eq:exp(p_squared)}) as 

\begin{multline}\label{eq:term20}
    4\int^\OM\int^\OM\int^\OM\int^\OM\int\int\int\int\la\AL(\M)^*\AL(\MB)\AL(\MBB)^*\AL(\MBBB)\x(\V-\M)\D\x(\VB-\MB)\x(\VBB-\MBB)\D\x(\VBBB-\MBBB)\ra d\M d\MB d\MBB d\MBBB d\V d\VB d\VBB d\VBBB\\ = 4\left\la\left|\int^\OM\int\AL(\M)\x(\V-\M)d\M d\V\right|^4\right\ra \\= 4\left\la\vphantom{\int}\right.\left|\vphantom{\int}\right.\PZ\AL\left(\int^\OM\x(\V)d\V + \int^\OM\int h(\M)\x(\V-\M)d\M d\V \right)+\frac{\PM\AL}{2}\left(\vphantom{\int}\right.\int^\OM\x(\V-\OM)d\V+\int^\OM\x(\V+\OM)d\V\\ + \int^\OM\int h(\M-\OM)\x(\V-\M)d\M d\V + \int^\OM\int h(\M+\OM)\x(\V-\M)d\M d\V \left.\vphantom{\int}\right)\left.\vphantom{\int}\right|^4\left.\vphantom{\int}\right\ra.
\end{multline}

In the expansion of Eq.~(\ref{eq:term20}), many terms vanish due to both the restricted domain of $h(\V)$ and the action of creation and annihilation operators on the vacuum. This allows Eq.~(\ref{eq:term20}) to be significantly simplified, leading to the result: 
\begin{multline}
   4\int^\OM\int^\OM\int^\OM\int^\OM\int\int\int\int\la\AL(\M)^*\AL(\MB)\AL(\MBB)^*\AL(\MBBB)\x(\V-\M)\D\x(\VB-\MB)\x(\VBB-\MBB)\D\x(\VBBB-\MBBB)\ra d\M d\MB d\MBB d\MBBB d\V d\VB d\VBB d\VBBB \\=(4\PZ^4+4\PZ^2\PM^2+\PM^4)\AL^4\Phi^2\left(\vphantom{\int}\right.\frac{1}{2}B^2 + 2B\int^\OM\int^\OM\la\R[h(\VB-\V)]\ra d\V d\VB \\+ \int^\OM\int^\OM\int^\OM\int^\OM\la\R[h(\VB-\V)h(\VBBB-\VBB)]\ra d\V d\VB d\VBB d\VBBB + \int^\OM\int^\OM\int^\OM\int^\OM\la h(\VB-\V)^*h(\VBBB-\VBB)\ra d\V d\VB d\VBB d\VBBB \\+ B\int^\OM\int^\OM\int\la h(\M)^*h(\M+\V-\VB)\ra d\M d\V d\VB + 2\int^\OM\int^\OM\int^\OM\int^\OM\int\la\R[h(\M)h(\M+\V-\VB)^*h(\VBBB-\VBB)]\ra d\M d\V d\VB d\VBB d\VBBB\\ + \frac{1}{2}\int^\OM\int^\OM\int^\OM\int^\OM\int\int\la h(\M)^*h(\M+\VB-\V)h(\MB)^*h(\MB+\VBBB-\VBB)\ra d\M d\MB d\V d\VB d\VBB d\VBBB\left.\vphantom{\int}\right)\\ + \frac{\PM^4}{8}\AL^4\Phi^2\left(\vphantom{\int}\right.\frac{1}{2}B^2 + 2B\int^\OM\int^\OM\la\R[h(\V+\VB-2\OM)]\ra d\V d\VB \\+ B\int^\OM\int^\OM\int\la\R[h(\M-\OM)h(\V+\VB-\M-\OM)]\ra d\M d\V d\VB+ 2\int^\OM\int^\OM\int^\OM\int^\OM\la h(\V+\VB-2\OM)^*h(\VBB+\VBBB-2\OM)\ra d\V d\VB d\VBB d\VBBB \\+ 2\int^\OM\int^\OM\int^\OM\int^\OM\int\la\R[h(\M-\OM)h(\V+\VB-2\OM)^*h(\VBB+\VBBB-\M-\OM)]\ra d\M d\V d\VB d\VBB d\VBBB\\ + \frac{1}{2}\int^\OM\int^\OM\int^\OM\int^\OM\int\int\la h(\M-\OM)^*h(\MB-\OM)h(\V+\VB-\M-\OM)^*h(\VBB+\VBBB-\M-\OM)\ra d\M d\MB d\V d\VB d\VBB d\VBBB \left.\vphantom{\int}\right).
\end{multline}
The summation of terms 21-24 of Eq.~(\ref{eq:exp(p_squared)}) can be written as
\begin{multline}
    2\int^\OM\int^\OM\int^\OM\int^\OM\int\int\int\int\left[\vphantom{\int}\right.\la\AL(\M)^*\AL(\MB)\x(\V-\M)\D\x(\VB-\MB)\hat{a}(\VBB-\MBB)\D\hat{a}(-\MBB)\hat{a}(-\MBBB)\D\hat{a}(\VBBB-\MBBB)\ra\\+ \la\AL(\M)^*\AL(\MB)\hat{a}(\VBB-\MBB)\D\hat{a}(-\MBB)\hat{a}(-\MBBB)\D\hat{a}(\VBBB-\MBBB)\x(\V-\M)\D\x(\VB-\MB)\ra \\+ \la\AL(\M)^*\AL(\MB)\x(\V-\M)\D\hat{a}(-\MBB)\D\hat{a}(\VB-\MBB)\hat{a}(\VBB-\MBBB)\D\hat{a}(-\MBBB)\x(\VBBB-\MB)\ra \\+ \la\AL(\M)^*\AL(\MB)\hat{a}(\VBB-\MBBB)\D\hat{a}(-\MBBB)\x(\VBBB-\MB)\x(\V-\M)\D\hat{a}(-\MBB)\D\hat{a}(\VB-\MBB)\ra \left.\vphantom{\int}\right]d\M d\MB d\MBB d\MBBB d\V d\VB d\VBB d\VBBB  \\=\left(\frac{5\Phi^3}{2}-2\Phi+\frac{1}{2\Phi}\right)\int^\OM\int^\OM\int^\OM\int^\OM\int\la\AL(\M)^*\AL(\M+\V+\VB-\VBB-\VBBB)\ra d\M d\V d\VB d\VBB d\VBBB\\=(2\PZ^2+\PM^2)\AL^2\left(\frac{5\Phi^3}{2}-2\Phi+\frac{1}{2\Phi}\right)\left(\vphantom{\int}\right.\frac{1}{2}B^3+\int^\OM\int^\OM\int^\OM\int^\OM\la\R[h(\V+\VB-\VBB-\VBBB)]\ra d\V d\VB d\VBB d\VBBB \\+ \frac{1}{2}\int^\OM\int^\OM\int^\OM\int^\OM\int\la h(\M)^*h(\M+\V+\VB-\VBB-\VBBB)\ra d\M d\V d\VB d\VBB d\VBBB\left.\vphantom{\int}\right).
\end{multline}
Similarly, we can combine terms 25 and 26 of Eq.~(\ref{eq:exp(p_squared)}) to give
\begin{multline}
    2\int^\OM\int^\OM\int^\OM\int^\OM\int\int\int\int\left[\vphantom{\int}\right.\la\AL(\M)^*\AL(\MB)^*\x(\V-\M)\D\hat{a}(-\MBB)\hat{a}(\VB-\MBB)\x(\VBB-\MB)\D\hat{a}(-\MBBB)\D\hat{a}(\VBBB-\MBBB)\ra\\+ \la\AL(\M)\AL(\MB)\hat{a}(\V-\MBB)\D\hat{a}(-\MBB)\x(\VB-\M)\hat{a}(\VBB-\MBBB)\D\hat{a}(-\MBBB)\x(\VBBB-\MB)\ra\left.\vphantom{\int}\right]d\M d\MB d\MBB d\MBBB d\V d\VB d\VBB d\VBBB \\=(2\PZ^2+\PM^2)\AL^2\left(\Phi^3-\Phi\right)\left(\vphantom{\int}\right.\frac{1}{2}B^3+\int^\OM\int^\OM\int^\OM\int^\OM\la\R[h(\V+\VB-\VBB-\VBBB)]\ra d\V d\VB d\VBB d\VBBB \\+ \frac{1}{2}\int^\OM\int^\OM\int^\OM\int^\OM\int\la \R[h(\M)h(\M+\V+\VB-\VBB-\VBBB)]\ra d\M d\V d\VB d\VBB d\VBBB\left.\vphantom{\int}\right).
\end{multline}
The final term of Eq.~(\ref{eq:exp(p_squared)}) gives the result:
\begin{multline}
    \int^\OM\int^\OM\int^\OM\int^\OM\int\int\int\int\la\A(\V-\M)\D\A(-\M)\A(-\MB)\D\A(\VB-\MB)\A(\VBB-\MBB)\D\A(-\MBB)\A(-\MBBB)\D\A(\VBBB-\MBBB)\ra d\M d\MB d\MBB d\MBBB d\V d\VB d\VBB d\VBBB \\=B^3\Lambda\left(\frac{7\Phi^4}{32}-\frac{3\Phi^2}{8}+\frac{7}{32\Phi^4}-\frac{3}{8\Phi^2}+\frac{5}{16}\right).
\end{multline}
By combining all the terms calculated for Eq.~(\ref{eq:exp(p_squared)}), we obtain the result for $\la p_\Omega^2\ra$:

\begin{multline}
    \la p_\Omega^2\ra = 4q^4R^2\Bigg[ \la |I_\OM|^4\ra + (4\PZ^2+2\PM^2)\AL^2\Phi\left(\vphantom{\int}\right.\la|I_\OM|^2\ra B + 2\int^\OM\int^\OM\la|I_\OM|^2\R[h(\V-\VB)]\ra d\V d\VB \\ + \int^\OM\int^\OM\int\la|I_\OM|^2h(\M)^*h(\M+\V-\VB)\ra d\M d\V d\VB\left.\vphantom{\int}\right) +\frac{\PM^2}{2}\AL^2\Phi\left(\vphantom{\int}\right.\la \R[I_\OM^{*2}]\ra B + 2\int^\OM\int^\OM\la \R[I_\OM^{*2}h(\V+\VB-2\OM)]\ra d\V d\VB \\+ \int^\OM\int^\OM\int\la \R[I_\OM^{*2}h(\M-\OM)h(\V+\VB-\M-\OM)]\ra d\M d\V d\VB \left.\vphantom{\int}\right) + \la|I_\OM|^2\ra\Lambda B\left(\frac{\Phi^2}{2}+\frac{1}{2\Phi^{2}}-1\right)  \\+ \PZ\PM\AL^2(4\Phi^2-2)\left(\vphantom{\int}\right.\la\R[I_\OM]\ra B^2 + 2\int^\OM\int^\OM\int^\OM\la\R[I_\OM]\R[h(\V+\VB-\VBB-\OM)]\ra d\V d\VB d\VBB \\+ \int^\OM\int^\OM\int^\OM\int\la\R[I_\OM]h(\M-\OM)h(\M+\V-\VB-\VBB)^*\ra d\M d\V d\VB d\VBB \left.\vphantom{\int}\right) \\ + 2\PZ\PM\AL^2\Phi^2\left(\vphantom{\int}\right.\la\R[I_\OM]\ra B^2 + 2\int^\OM\int^\OM\int^\OM\la\R[I_\OM^* h(\V+\VB-\VBB-\OM)]\ra d\V d\VB d\VBB \\+ \int^\OM\int^\OM\int^\OM\int\la\R[I_\OM^*h(\M-\OM)h(\M+\V-\VB-\VBB)]\ra d\M d\V d\VB d\VBB \left.\vphantom{\int}\right) \\ + (4\PZ^4+4\PZ^2\PM^2+\PM^4)\AL^4\Phi^2\left(\vphantom{\int}\right.\frac{1}{2}B^2 + 2B\int^\OM\int^\OM\la\R[h(\VB-\V)]\ra d\V d\VB \\+ \int^\OM\int^\OM\int^\OM\int^\OM\la\R[h(\VB-\V)h(\VBBB-\VBB)]\ra d\V d\VB d\VBB d\VBBB + \int^\OM\int^\OM\int^\OM\int^\OM\la h(\VB-\V)^*h(\VBBB-\VBB)\ra d\V d\VB d\VBB d\VBBB \\+ B\int^\OM\int^\OM\int\la h(\M)^*h(\M+\V-\VB)\ra d\M d\V d\VB + 2\int^\OM\int^\OM\int^\OM\int^\OM\int\la\R[h(\M)h(\M+\V-\VB)^*h(\VBBB-\VBB)]\ra d\M d\V d\VB d\VBB d\VBBB\\ + \frac{1}{2}\int^\OM\int^\OM\int^\OM\int^\OM\int\int\la h(\M)^*h(\M+\VB-\V)h(\MB)^*h(\MB+\VBBB-\VBB)\ra d\M d\MB d\V d\VB d\VBB d\VBBB\left.\vphantom{\int}\right)\\ + \frac{\PM^4}{8}\AL^4\Phi^2\left(\vphantom{\int}\right.\frac{1}{2}B^2 + 2B\int^\OM\int^\OM\la\R[h(\V+\VB-2\OM)]\ra d\V d\VB + B\int^\OM\int^\OM\int\la\R[h(\M-\OM)h(\V+\VB-\M-\OM)]\ra d\M d\V d\VB\\+ 2\int^\OM\int^\OM\int^\OM\int^\OM\la h(\V+\VB-2\OM)^*h(\VBB+\VBBB-2\OM)\ra d\V d\VB d\VBB d\VBBB \\+ 2\int^\OM\int^\OM\int^\OM\int^\OM\int\la\R[h(\M-\OM)h(\V+\VB-2\OM)^*h(\VBB+\VBBB-\M-\OM)]\ra d\M d\V d\VB d\VBB d\VBBB\\ + \frac{1}{2}\int^\OM\int^\OM\int^\OM\int^\OM\int\int\la h(\M-\OM)^*h(\MB-\OM)h(\V+\VB-\M-\OM)^*h(\VBB+\VBBB-\M-\OM)\ra d\M d\MB d\V d\VB d\VBB d\VBBB \left.\vphantom{\int}\right) \\ + (2\PZ^2+\PM^2)\AL^2\left(\frac{5\Phi^3}{2}-2\Phi+\frac{1}{2\Phi}\right)\left(\vphantom{\int}\right.\frac{1}{2}B^3+\int^\OM\int^\OM\int^\OM\int^\OM\la\R[h(\V+\VB-\VBB-\VBBB)]\ra d\V d\VB d\VBB d\VBBB \\+ \frac{1}{2}\int^\OM\int^\OM\int^\OM\int^\OM\int\la h(\M)^*h(\M+\V+\VB-\VBB-\VBBB)\ra d\M d\V d\VB d\VBB d\VBBB\left.\vphantom{\int}\right) \\+ (2\PZ^2+\PM^2)\AL^2\left(\Phi^3-\Phi\right)\left(\vphantom{\int}\right.\frac{1}{2}B^3+\int^\OM\int^\OM\int^\OM\int^\OM\la\R[h(\V+\VB-\VBB-\VBBB)]\ra d\V d\VB d\VBB d\VBBB \\+ \frac{1}{2}\int^\OM\int^\OM\int^\OM\int^\OM\int\la \R[h(\M)h(\M+\V+\VB-\VBB-\VBBB)]\ra d\M d\V d\VB d\VBB d\VBBB\left.\vphantom{\int}\right) \\+ B^3\Lambda\left(\frac{7\Phi^4}{32}-\frac{3\Phi^2}{8}+\frac{7}{32\Phi^4}-\frac{3}{8\Phi^2}+\frac{5}{16}\right)\Bigg].
\end{multline}

In the expression for $\text{Var}(p_\Omega)=\la p_\OM^2\ra-\la p_\OM\ra^2$, there is significant cancellation between $\la p_\OM^2\ra$ and $\la p_\OM\ra^2$, and by taking the leading remaining terms we find that 
\begin{multline}
    \text{Var}(p_\OM)=\la p_\OM^2\ra-\la p_\OM\ra^2\approx4q^4R^2\left[\la|I_\OM|^4\ra-\la|I_\OM|^2\ra^2 + 2\la|I_\OM|^2\ra\PZ^2\AL^2\Phi B\right]\\ \approx4q^4R^2\left[\vphantom{\int}\right.16\PZ^4\PM^4\AL^8\left(\int^\OM\int^\OM\la\R[h(\V-\OM)]\R[h(\VB-\OM)]\ra d\V d\VB-\int^\OM\int^\OM\la\R[h(\V-\OM)]\ra\la\R[h(\VB-\OM)]\ra d\V d\VB\right) \\+ 4\PZ^2\PM^2\AL^4\left(\int^\OM\int^\OM\la\R[\n(\V)]\R[\n(\VB)]\ra d\V d\VB-\int^\OM\int^\OM\la\R[\n(\V)]\ra\la\R[\n(\VB)]\ra d\V d\VB\right) +2\Phi B\PZ^4\PM^2\AL^6\left.\vphantom{\int}\right].
\end{multline}
By associating $\mathcal{H}=\int^\OM h(\V-\OM)d\V=\int_{-B/2}^{B/2}h(\V)d\V$ as the relative amplitude of the classical optical noise in the DC component, $\mathcal{N}=\int^\OM\n(\V)d\V$ as the amplitude of the electronic noise in the $\pm B/2$ frequency interval around $\OM$, and substituting $\PZ\approx1$, $\PM=\delta_m/2$ and $i_0\approx q\eta\AL_0^2$, we find that for $M$ spectral averages,
\begin{equation}{\label{eq:var_p_approx}}
    \text{Var}(p_\Omega)\approx \frac{R^2}{M}\left[\vphantom{\frac{R^2}{M}}\right.2q\delta_m^2i_0^3\Phi B+ 4\delta_m^4i_0^4\text{Var}(\mathfrak{R}[\mathcal{H}])\\+4q^2\delta_m^2i_0^2\text{Var}(\mathfrak{R}[\mathcal{N}])\vphantom{R^2}\left.\vphantom{\frac{R^2}{M}}\right].
\end{equation}
From Eq.~(\ref{eq:var_p_approx}), an improvement in precision beyond the quantum noise limit may be obtained in the case that squeezing ($\Phi<1$) provides a significant reduction in $\text{Var}(p_\OM)$.

\end{document}